# Microwave assisted magnetization reversal in single domain nanoelements[1]

H. T. Nembach[1], H. Bauer[1], J. M. Shaw[1], M. L. Schneider[2] and T.J. Silva[1]

1. Electromagnetics Division, National Institute of Standards and Technology, Boulder, CO 80305
2. Department of Physics and Astronomy, University of Montana, Missoula, MO 59812

We studied the microwave assisted magnetic reversal of 65 nm × 71 nm elliptical $Ni_{80}Fe_{20}$ nanomagnets. Hysteresis curves were measured by magneto-optical Kerr effect for a range of microwave frequencies and amplitudes. The coercive field $H_c$ was reduced by 26% for an rf field of $0.08 \cdot H_c$ when the microwave frequency coincided with the minimum of the experimentally determined ferromagnetic resonance frequency with varying dc field. The experimental results for the fractional reduction in $H_c$ with rf field amplitude are in good agreement with numerical simulations for an array of interacting macrospins with a physically realistic shape anisotropy distribution.

As bit densities in magnetic hard drives approach 0.16 Tbit/cm$^2$ ($10^{12}$ bits per sq. in.), the anisotropy of the media must be significantly increased in order to maintain thermal stability. Such high values of anisotropy will require write fields that are beyond the fundamental limit imposed by the saturation magnetization $M_S$ of the write head materials. Energy assisted magnetic recording is one solution permitting bits in materials with high crystalline anisotropy to be written at fields easily attainable by existing write head technology. One form of energy assisted magnetic recording being pursued is microwave assisted magnetic recording (MAMR) [1]. In the case of MAMR, an rf field is tuned to the ferromagnetic resonance frequency of the recording medium, while a quasi-dc field is also applied, where the quasi-dc field is smaller than what the switching field would be in the absence of microwaves. Resonant magnetic precession drives the magnetization over the energy barrier imposed by anisotropy, provided that the rf field amplitude is sufficiently large.

MAMR has been studied experimentally in Co nanoparticles [2], Permalloy elements [3], [4], [5] and magnetic tunnel junctions [6]. So far, there have been no reports of MAMR experiments for single domain magnetic elements with dimensions that approach those that are required for either bit patterned recording media or continuous media with exchange decoupled grains.

In this letter, we report experimental and numerical results for the dependence of the switching field on the rf field amplitude and frequency in the case of $Ni_{80}Fe_{20}$ nanomagnets. Although $Ni_{80}Fe_{20}$ is too soft magnetically to serve as an effective recording medium, it allows for a demonstration of MAMR with nanomagnets that are sufficiently small to be considered approximate macrospins.

A 10 nm thick film was grown by dc magnetron sputtering on a sapphire substrate with a 3 nm thick Ta seed layer and capped with 5 nm $Si_3N_4$ to prevent oxidation. An array with 65 nm × 71 nm elliptical nanodots with 100 nm center-to-center spacing was patterned by electron beam lithography. Details of the sample preparation are given in Ref. 7.

Hysteresis curves were measured by use of the transverse magneto-optical Kerr effect, whereby the Kerr signal is proportional to the magnetization component perpendicular to the optical plane of incidence. An external dc field used to switch the elements was also aligned perpendicular to the plane of incidence. The long axes of the magnetic elements were aligned parallel to the external dc field. The microwave field was oriented perpendicular to the external field. A continuous wave laser with a wavelength of 532 nm was focused onto the array at a 45 degree angle of incidence. The laser power at the sample was 5 mW, and the laser spot size 8 μm, such that approximately 5000 nanomagnets were illuminated. A cw microwave field was applied while the external magnetic field was swept. Microwaves were applied to the sample by a coplanar waveguide with a center conductor width of 150 μm

Fig.1 shows a color contour plot of the coercive field as a function of the microwave frequency and amplitude. At 2.15 GHz, the coercive field $\mu_0 H_c$ = 5.2 mT was reduced by 1.4 mT (a 26 % reduction) with an rf field amplitude of only 0.4 mT. Fig. 2(a) shows hysteresis curves with and without an rf field of 2.15 GHz and an amplitude of 0.4 mT. The reduction in the coercive field can be seen clearly. The nucleation field shows a smaller reduction. The dependence of the nucleation field and the coercive field on the rf field amplitude for a frequency of 2.15 GHz is shown in Fig. 2(c). The maximum reduction of the nucleation field is 0.5 mT. This is smaller than the reduction of the coercive field; i.e., the hysteresis curves become steeper with increasing microwave amplitude.

The ferromagnetic resonance (FMR) frequency as a function of dc applied field was determined using frequency-resolved magneto-optical Kerr effect for the same sample [7]. Prior to each measurement, the sample was saturated at a positive magnetic field. For these measurements, the rf-field was 0.03 mT, thus excluding any influence of the rf field on the reversal behavior. (Exclusion of laser heating effects are discussed later in this Letter.) The measured FMR frequencies are shown in Fig.2(b). For clarity, we plotted the branch for increasing applied field (red triangles) by inverting the measured curve for decreasing applied field (black circles). The magnetic field region between the nucleation and saturation fields is indicated by the dotted vertical lines. The inset in Fig. 2(b) shows a resonance spectrum for an applied dc magnetic field of $\mu_0 H$ = -4.7 mT. For microwave assisted magnetic reversal to occur, the magnetization must be resonantly excited in the field range between the nucleation of the switching process and $H_c$. The FMR frequencies in this region range from 2.1 GHz to 2.5 GHz. The minimum of $H_c$ occurred for an rf frequency of 2.15 GHz, which is close to the minimum of the FMR frequency for varying dc field. While $H_c$ is reduced with increasing rf amplitude, the frequency at which $H_c$ is minimized remains approximately constant with varying rf amplitude. This is understandable because the dependence of the FMR frequency on applied field (60 MHz/mT) is too shallow to cause a substantial change in the resonance frequency at $H_c$ for the range of rf amplitudes employed in this study.

To exclude laser heating effects, hysteresis curves without applied microwaves were measured for a broad range of laser powers. There were no significant changes in the coercive field for a laser power of less than 10 mW.

Microwave heating effects can be excluded from simple absorption calculations. The microwave power absorption per unit area can be calculated in a similar way as that in Ref. 4 and Ref. 8, and is given by $P = \pi \mu_0 f \chi_{xx} h^2 d$, where $\chi_{xx}$ is the imaginary part of the *xx*-component of the Polder susceptibility tensor for a resonant microwave frequency $f$ = 2.15 GHz, and $d$ is the thickness of the magnetic layer. For the sample parameters found in Ref. 7, we calculate $\chi_{xx} = 1.2 \times 10^3$ at the measured resonance field. If we assume only heat conduction perpendicular to the substrate surface and that the backside of the substrate is at room temperature, Fourier's law for heat conduction yields a nanomagnet temperature rise of only $\Delta T = \frac{Pt}{\lambda} = 0.04$ K, for a substrate thickness $t$ = 120 μm and the heat conductivity for sapphire $\lambda$ = 35 W·m$^{-1}$·K$^{-1}$ [9].

Micromagnetic simulations at $T$ = 0 K were performed by use of OOMMF [10] for a single element. Simulation parameters were: saturation magnetization $M_s$ = 800 k·Am$^{-1}$, exchange constant $A$ = 1.3 10$^{-11}$ Jm$^{-1}$ and damping constant $\alpha$ = 0.008 [7]. The mode with the lowest resonance frequency was localized at the ends of the ellipsoidal element [11]. The resonance frequency agrees well with the experimental value in Ref. 7. The magnetic reversal of a single nanomagnet with an applied microwave field of $\mu_0 H_{rf}$ = 0.3 mT was then simulated. The result is shown in Fig. 3(a). The dependence of the $H_c$ on the microwave frequency is asymmetric with respect to the frequency: on the high frequency side, $H_c$ has a sharp dependence on frequency, unlike what was observed in our experiments.

We also used a custom simulation program to model microwave assisted reversal in a 20×20 nanomagnet array (also at $T$ = 0) that incorporated dipolar interactions between the nanomagnets, where we approximated each nanomagnet as a macrospin with point dipole magnetostatic fields. For nearest neighbors, we introduced a correction factor to account for errors in the point dipole approximation. This correction factor was calculated by dividing the nanomagnets into $1nm^2$ cells and determining the dipolar field in each cell of the neighboring nanomagnet via numerical integration. Moreover, the magnetic moment of each nanomagnet was reduced by 0.8 to account for the inhomogenous magnetization during reversal. We used a Gaussian size distribution with a standard deviation of 1.5 nm, consistent with previous FMR linewidth measurements in the same sample [7]. The external applied field was oriented 5 degrees from the easy axis.

Hysteresis curves were determined from the average magnetization of the array. Because the behavior of the outermost nanomagnets differed significantly from that of the interior ones, the nanomagnets in the outermost three rows and columns were neglected. Moreover, we performed anywhere from 40 to 60 simulations, each with a different set of random nanomagnet dimensions, and averaged the results to construct each hysteresis curve. The inset in Fig. 2(a) shows a simulated hysteresis curve without an applied microwave field. In Fig. 3(b), we show the dependence of the coercive field on the microwave frequency for an rf field amplitude of 0.35 mT. Relative to the micromagnetic simulations for a single nanomagnet, the reduction in $H_c$ for the dipole array is more symmetric with respect to frequency, the fractional reduction in $H_c$ is greater, and the range of microwave frequencies that reduce $H_c$ is broader.

Fig.4 in a color contour plot of our simulation results with varying rf field amplitude and frequency, which are qualitatively similar to the experimental results in Fig. 1. $H_c$ is reduced from $\mu_0 H_c$ = 6.4 mT to 5.6 mT with an rf field amplitude of 0.4 mT at $f$ = 1.7 GHz. However, the modeled values of $H_c$ are 1.3 times higher than the experimental values. The discrepancy can be explained in terms of thermal fluctuations that assist the switching process. The measured $H_c$ at room temperature agrees well with the prediction of conventional Arrhenius-Néel theory for a single domain nanomagnet [12]. The larger value of $H_c$ for the simulations relative to $T$ = 0 simulations also explains the reduction of the frequency where $H_c$ is minimized relative to the experimental data: the resonance frequency decreases with increasing field when the dc field is applied antiparallel to the magnetization, resulting in a suppressed resonance frequency near $H_c$ for the simulations at $T$ = 0.

In summary, we demonstrated that $H_c$ in single domain $Ni_{80}Fe_{20}$ nanomagnets can be reduced by 26% through application of an rf field with an amplitude of only 8% of $H_c$. For the modest microwave amplitudes employed in these experiments, the rf frequency at which $H_c$ is minimized is coincident with the minimum value of the FMR frequency with varying applied dc field. Dipole array simulations that include the effects of size distribution correctly reproduce the fractional variation of $H_c$ with varying rf frequency, though the values for $H_c$ and the frequency that minimizes $H_c$ differ by 20% from the experimental results due to the omission of thermal fluctuations from the numerical models.

[1] J.-G. Zhu, IEEE Trans. Magn. 44, 125 (2008)

[2] C. Thirion, W. Wernsdorfer and D. Mailly, Nat. Mater. 2, 524, (2003)

[3] H. T. Nembach, P. Martin Pimentel, S.J. Hermsdoerfer, B. Leven, B. Hillebrands and S.O. Demokritov, Appl. Phys. Lett., 90, 062503 (2007)

[4] G. Woltersdorf, C.H. Back, Phys. Rev. Lett., 99, 227207 (2007)

[5] Y. Nozaki, K. Tateishi, S. Taharazako, M. Ohta, S. Yoshimura and K. Matsuyama, Appl. Phys. Lett., 91, 122505 (2007)

[6] T. Moriyama, R. Cao, J.Q. Xiao, J. Lu, X.R. Wang, Q. Wen and H.W. Zhang, Appl. Phys. Lett., 90, 152503 (2007)

[7] M.L. Schneider, J.M. Shaw, A.B. Kos, Th. Gerrits, T.J. Silva and R.D. McMichael, J. Appl. Phys. 102, 103909 (2007)

[8] R. Meckenstrock, Rev. Sci. Instrum., 79, 041101 (2008)

[9] CRC Handbook of Chemistry and Physics, Editor D.R. Lide, 89$^{th}$ Edition, page 12-206, CRC Press

[10] M.J. Donahue and D.G. Porter., Internal Report NISTIR 6376, National Institute of Standards and Technology (1999)

[11] J.M.Shaw, M.L. Schneider, T.J. Silva and R.D. McMichael, Phys. Rev B, 79, 184404 (2009)

[12] L. Néel, Ann. Geophys. **5**, 99 (1949).

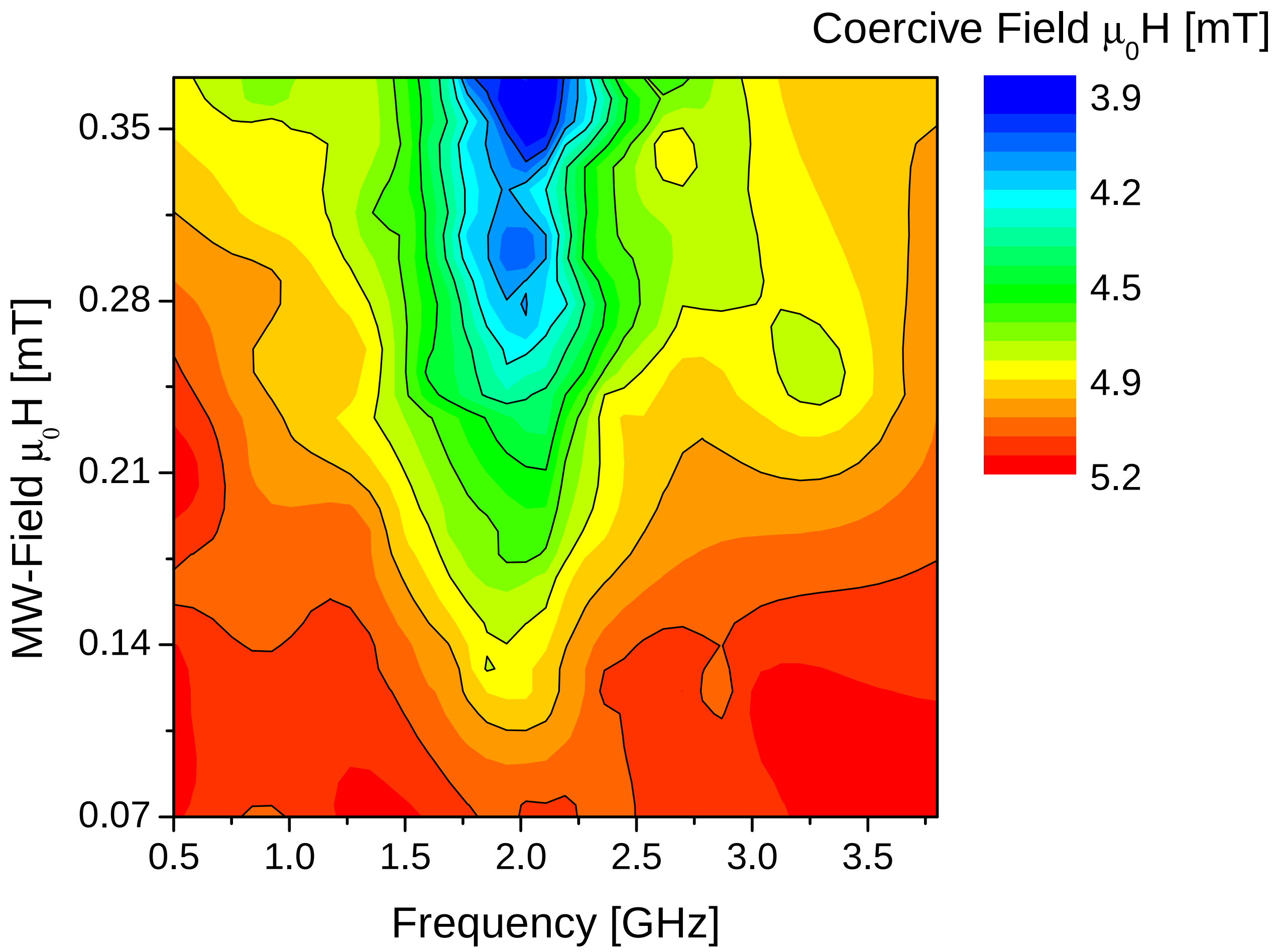

Figure 1: Contour plot of the coercive field as functions of rf field and frequency. The coercive field is is strongly reduced at microwave frequencies in the region around 2 GHz and decreases as the microwave field amplitude is increased.

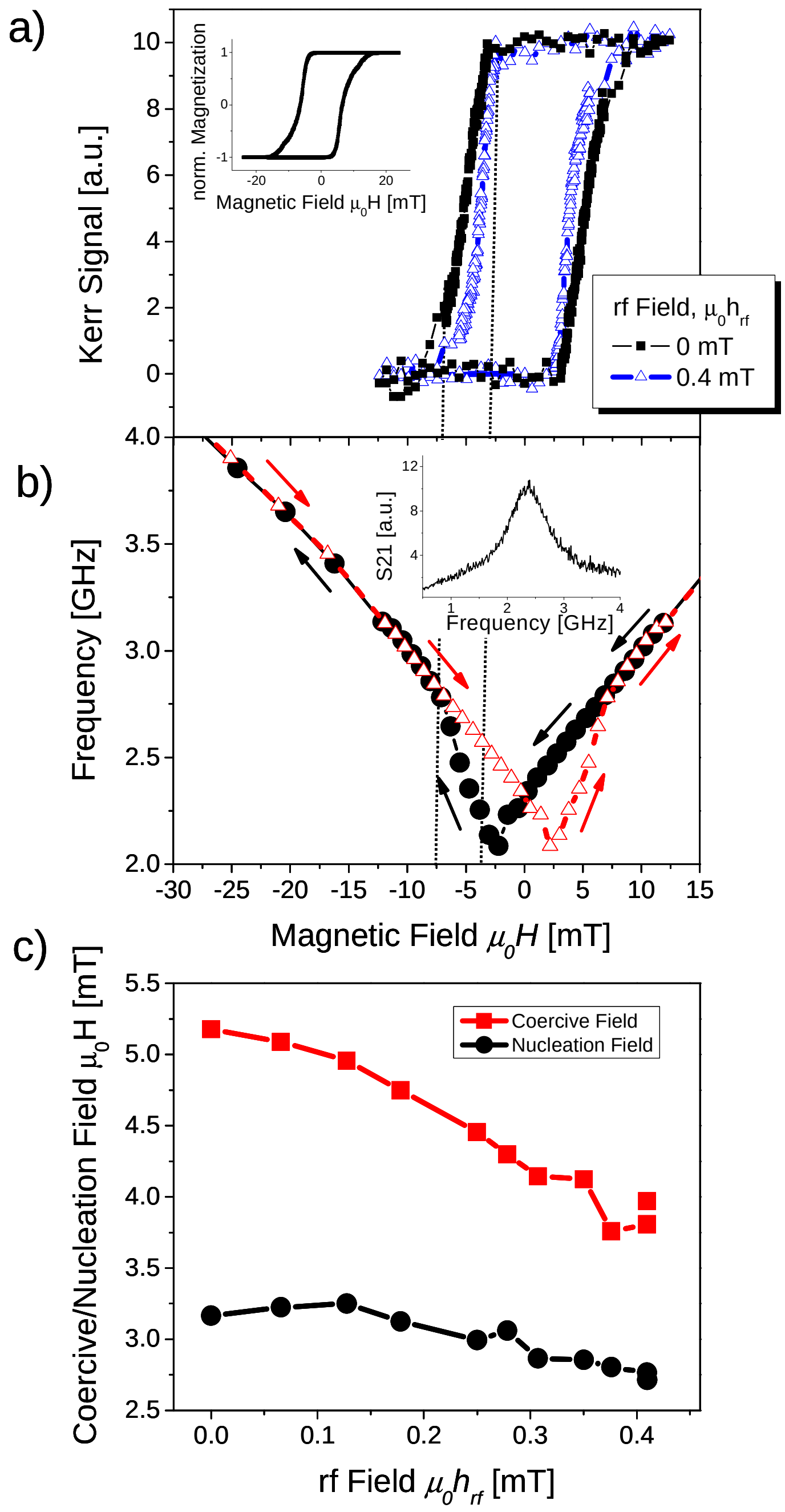

Figure 2. (a) Hysteresis curves with and without an applied microwave field of 0.4 mT and a frequency of 2.15 GHz. Application of the microwave field reduces the coercive field. The inset shows a simulated hysteresis curve for a 20×20 array without applied microwaves. (b) FMR frequency versus applied magnetic field. The curve for increasing applied field (red triangles) was obtained by inverting the measured curve for decreasing applied field (black circles). The sample was saturated with a positive magnetic field prior to each measurement. The inset shows a resonance curve for an applied field of -4.7 mT. The vertical dashed lines in (a) and (b) indicate the fields at which reversal begins and ends for an rf field of 0.4 mT. (c) Dependence of the coercive (red squares) and the nucleation (black circles) field on the amplitude of the rf field for a frequency of 2.15 GHz.

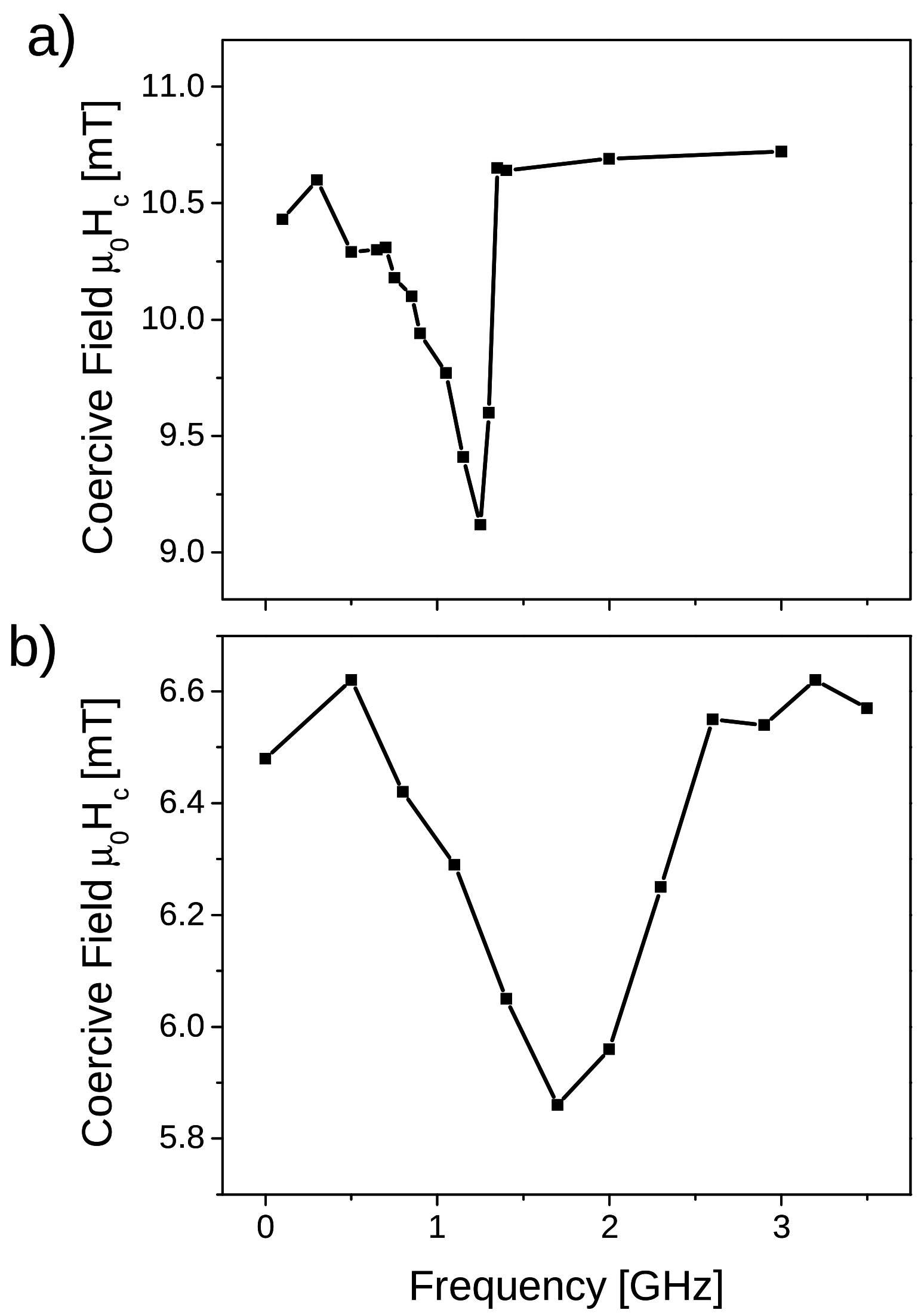

Figure 3: The coercive field versus microwave frequency is shown for micromagnetic simulations for a single magnetic element (a) and for the dipole approximation of a 20 × 20 nanodot array (b). The microwave amplitudes were respectively 0.3 mT and 0.35 mT.

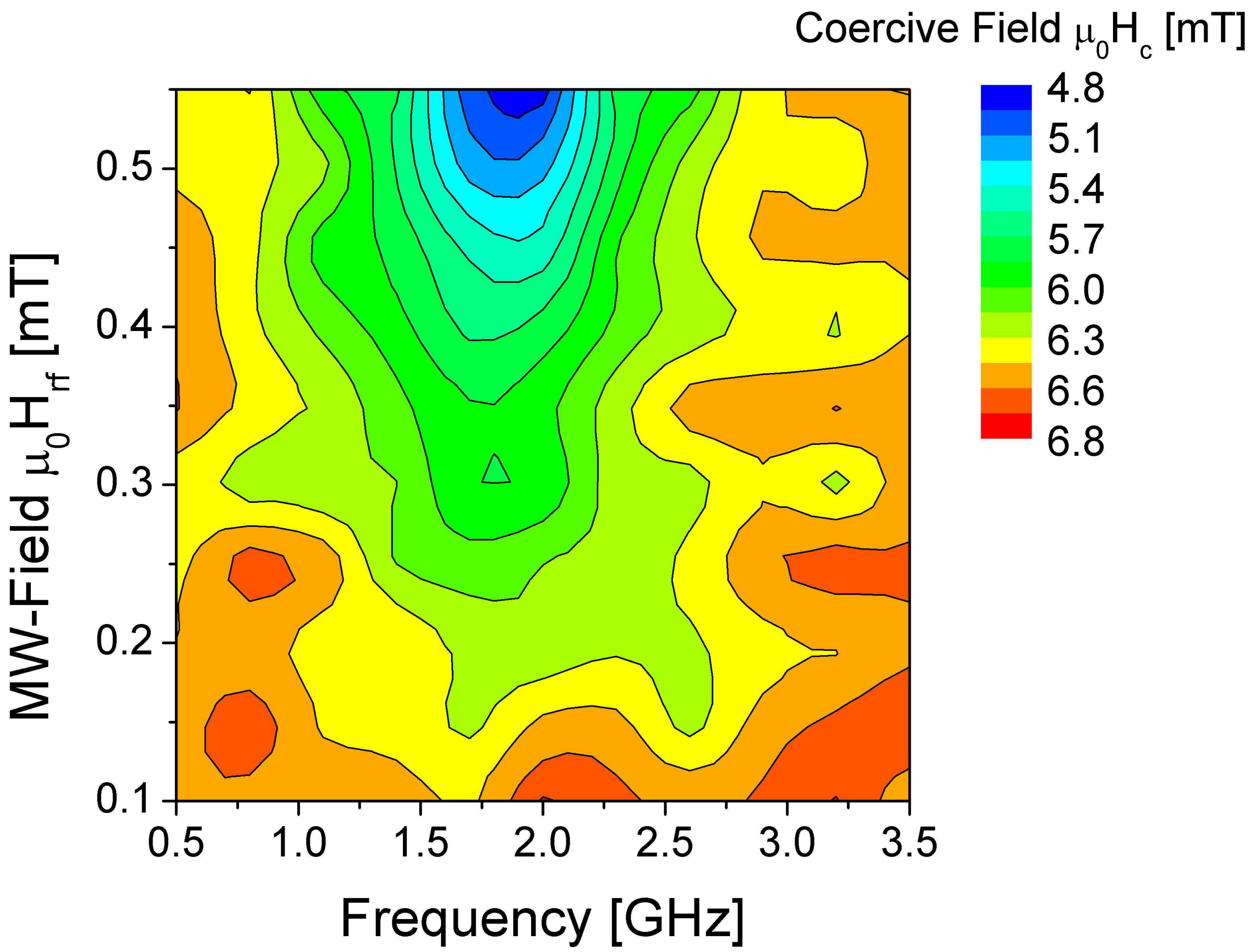

Figure 4: The coercive field was determined by simulating the magnetic reversal of a 20 × 20 nanodot array with an applied microwave field for varying frequency and amplitude.